\documentstyle[12pt]{article}
\begin{document}
\title{Determination of polarised parton distributions in the nucleon - next 
to leading order QCD analysis}
\author{Stanis\l aw  Tatur \\
Nicolaus Copernicus Astronomical Center,\\ Polish Academy of Sciences,\\
Bartycka 18, 00-716 Warsaw, Poland. \\ \\
\and
Jan Bartelski, Miros\l aw Kurzela\\
Institute of Theoretical Physics, Warsaw University,\\
Ho$\dot{z}$a 69, 00-681 Warsaw, Poland. \\}
\date{}
\maketitle
\vspace{1cm}
\begin{abstract}
\noindent
We have made next to leading order QCD fit to the deep inelastic spin
asymmetries on nucleons and we determined polarised quark and gluon 
densities. The functional form for such distributions was inspired by the 
Martin, Roberts and Stirling fit for unpolarised case. In addition to usually 
used data points (averaged over $x$ and $Q^2$) we have also considered 
the sample containing points with similar $x$ and different $Q^2$. 
It seems that splitting of quark densities into valence and sea 
contribution is strongly model dependent and only their sum (i.e., 
$\Delta u$ and $\Delta d$) can be precisely determined from the data. 
Integrated polarised gluon contribution, contrary to some expectations, is 
relatively small and the sign of it depends on the fact which sample of 
data points is used.
\end{abstract}

\newpage
The final analysis of data on polarised deep inelastic scattering taken in 
E143 experiment at SLAC and SMC at CERN on proton and deuteron targets were 
published recently \cite{e143new,SMCnew}. Together with the older data from 
SLAC \cite{e80,e130,e142n,e143p,e143d,e154n,e143qd}, CERN 
\cite{emc,SMCd,SMCp,SMCqp,SMCqd} and DESY \cite{hermes} one has quite a lot 
data on spin asymmetries. However, the newest data has 
smaller statistical errors and hence dominates in $\chi^2$ fits. Study of 
polarised deep inelastic scattering were for the first time performed by 
SLAC-Yale group \cite{e80,e130}. After the results of EMC group from CERN 
\cite{emc} leading to the so called spin crisis there was enormous 
interest in studying polarised structure functions. It was suggested 
\cite{dg} that polarised gluons may be responsible for the little spin 
carried by quarks. Experimental groups have measured spin 
asymmetries on proton, neutron ($^3 He$) and deuteron targets. On the other 
hand after calculation of two loop polarised splitting functions \cite{ew} 
several next to leading (NLO) order QCD 
analysis were performed \cite{grv,alt0,smcth,inni} making fits to the 
actually existing data and trying to determine polarised parton distributions. 
Determination of polarised gluon distribution was particularly 
interesting in this context. The aim of this 
paper is similar. We want to perform next to leading order fit to the data 
taking into account the recently published data from the new analysis (SLAC 
E143 and CERN SMC experiments). We will divide 
the data into two groups. Many experimental groups published data 
\cite{e143new,SMCnew} sets for the close values of $x$ and different $Q^2$ 
in addition to the "averaged" data where one averages over $x$ and $Q^2$ 
(the errors are smaller and $Q^2$ dependence is smeared out). In principle 
when we take into account $Q^2$ evolution of polarised and unpolarised 
functions (in NLO analysis) the first group of data points, i.e. non 
averaged, should give better fit. In most of the fits to experimental data 
only second group of data namely with averaged $x$ and $Q^2$ dependence was 
used. We will make fits to the 
both sets of data (the first group contains 374 points and the second 130 
points) and compare them with the fits without $Q^2$ evolution taken into 
account (in other words assuming that asymmetries do not depend on $Q^2$). 
One should add that many experimental groups have not succeeded in finding 
$Q^2$ dependence for approximately the same value of $x$ and different $Q^2$ 
\cite{e143qd,SMCqp,SMCqd}. The open 
question is if we really see the $Q^2$ evolution of structure functions from 
the existing experimental data. In our analysis we limit ourselves, as one 
usually does, to the data with $Q^2 \geq$ 1 $\mbox{GeV}^2$. As was already 
mentioned in our earlier papers \cite{bt1} and was later stressed by other 
authors \cite{grv} making a fit to spin asymmetries and not directly to $g_1 
(x,Q^2)$ we expect that the higher twist contributions are probably less 
important in such 
case. Experiments on unpolarised DIS provide information on the unpolarised 
quark densities $q(x,Q^2)$ and $G(x,Q^2)$ inside the nucleon. These densities 
can be expressed in term of $q^{\pm}(x,Q^2)$ and $G^{\pm}(x,Q^2)$, i.e. 
densities of quarks and gluons with helicity along or opposite to the 
helicity of the parent nucleon. The unpolarised quark densities are given by 
the sum of $q^{+}$, $q^{-}$ and $G^{+}$  , $G^{-}$, namely: 
\begin{equation}
q = q^{+}+q^{-},\hspace*{1.5cm} G = G^{+}+G^{-}.
\end{equation}

The polarised DIS experiments give also information about the so called 
polarised parton density, the difference of $q^{+}$, $q^{-}$ and $G^{+}$, 
$G^{-}$:
\begin{equation}
\Delta q = q^{+}-q^{-},\hspace*{1.5cm} \Delta G = G^{+}-G^{-}.
\end{equation}

We will actually try to determine $q^{\pm} (x,Q^2)$ and $G^{\pm} (x,Q^2)$, 
in other words, we will have in some sense a simultaneous fit to unpolarised 
and polarised data. In 
principle the asymptotic x behaviour of $q^{+}$ and $q^{-}$ will be taken 
from the unpolarised data. We will use fits of MRS (called R2) \cite{mrsn} 
taking into account the behaviour at small $x$ of 
quarks and gluon distributions obtained from experiments in Hera. It is of 
course very restrictive assumption that $\Delta q$ and $\Delta G$ have the 
same behaviour (when the integrals of $\Delta q$ and $\Delta G$ exist) as 
$q$ and $G$. On the other hand the small $x$ behaviour of unpolarised 
structure functions is determined from the $x$ values of Hera much smaller 
than in the polarised case. We will also consider the integrals over the 
region measured in the experiments with polarised particles with the hope 
that in this case the behaviour of $q^{\pm}$ and $G^{\pm}$ could be more 
plausible. The values of integrals in the whole region ($0 \leq x \leq 1$) 
involving asymptotic behaviour taken from the unpolarised structure functions 
may be not as reliable as for the measured region. But it is an alternative 
to use of Regge type behaviour. 

It is known \cite{grv} that the behaviour of the quark and gluon 
distributions in small $x$ region is extremely important in extrapolation of 
calculation of integrals over whole $ 0 \leq x \leq 1$ range. It could 
happen that in $\Delta q = q^{+} - q^{-}$ (when we assume that $q^{+}$, 
$q^{-}$ and $q = q^{+} + q^{-}$ have similar $x$ dependence) most singular 
$x$ terms cancel (that is especially important in case of $d$ valence quark, 
sea and gluon where the $x$ behaviour is relatively singular). We will see 
how such description infers the fits and calculated parameters. With the 
less singular  distributions for $\Delta d_v$, $\Delta M$ (total sea 
polarisation) and $\Delta G$ (gluon polarisation) there is no strong
 dependence of calculated quantities in an unmeasured region but the fits 
have higher $\chi^2$. One of the main tasks of considering NLO evolution in 
$Q^2$ is the determination of the gluon contribution $\Delta G$. In 
$\overline{MS}$ scheme $\Delta G(x,Q^2)$ comes in through the higher order
 corrections. In our fits we obtain $\Delta G$ relatively small contrary to 
some expectations. In addition 
when we use the averaged sample of data $\Delta G$ contribution is opposite 
to that in a set without averaging over $x$ and $Q^2$. In our fits the 
averaging over $x$ and $Q^2$ (that have nothing to do with physics) changes 
the sign of the gluon contribution with rather small changes in quark 
region. It means that gluon contribution is extremely sensitive and can not 
be reliably determined. 

Let us start with the formulas for unpolarised quark parton distributions 
given (at $ Q^{2}=1\, {\rm GeV^{2}}$) in one of the recent fits performed by 
Martin, Roberts and Stirling \cite{mrsn}. In this fit 
$\Lambda^{n_{f}=4}_{\overline{MS}}=0.344$ $\mbox{GeV}$ and 
$\alpha_s(M^2_Z)=0.120$.  We have for the valence quarks:
\begin{eqnarray}
u_{v}(x)&=&2.251 x^{-0.39}(1-x)^{3.54}(1-0.98\sqrt{x}+6.51x),
\nonumber \\
d_{v}(x)&=&0.114 x^{-0.76}(1-x)^{4.21}(1+7.37\sqrt{x}+29.9x) ,
\end{eqnarray}

\noindent and for the antiquarks from the sea:
\begin{eqnarray}
2\bar{u} (x)&=&0.392M(x)-\delta (x), \nonumber \\
2\bar{d} (x)&=&0.392M(x)+\delta (x), \nonumber \\
2\bar{s} (x)&=&0.196M(x), \\
2\bar{c} (x)&=&0.020M(x). \nonumber
\end{eqnarray}

\noindent In eq.(4) the singlet contribution $M=2[\bar{u}+\bar{d}
+\bar{s}+\bar{c}$] is:
\begin{equation}
M(x)=0.37 x^{-1.15}(1-x)^{8.27}(1+1.13\sqrt{x}+14.4x),
\end{equation}

\noindent whereas the isovector $\delta=\bar{d}-\bar{u}$ part:
\begin{equation}
\delta (x)=0.036 x^{-1.15}(1-x)^{8.27}(1+64.9x).
\end{equation}

\noindent For the unpolarised gluon distribution we get:
\begin{equation}
G(x)=14.4 x^{-0.49}(1-x)^{5.51}(1-4.20\sqrt{x}+6.47x).
\end{equation}

We assume, in an analogy to the unpolarised case, that the polarised
quark distributions are of the form: 
$x^{\alpha}(1-x)^{\beta}P_{2}( \sqrt{x})$, where $P_{2}(\sqrt{x})$ is a 
second order polynomial in $\sqrt{x}$ and the asymptotic behaviour for 
$x \rightarrow 0$ and $x \rightarrow 1$ (i.e. the values of $\alpha$ and 
$\beta$) are the same (except for $\Delta M$, see a discussion below) as 
in unpolarised case. Our idea is to split the numerical constants 
(coefficients of $P_{2}$ polynomial) in eqs.(3, 5, 6 and 7) in two parts in 
such a manner that the distributions are positive defined. Our expressions 
for $\Delta q(x) = q^{+}(x)-q^{-}(x)$ ($q(x) = q^{+}(x)+q^{-}(x)$) are:
\begin{eqnarray}
\Delta u_{v}(x)&=&x^{-0.39}(1-x)^{3.54}(a_{1}+a_{2}\sqrt
{x}+a_{3}x), \nonumber \\
\Delta
d_{v}(x)&=&x^{-0.76}(1-x)^{4.21}(b_{1}+b_{2}\sqrt{x}+b_{3}x), \nonumber \\
\Delta M(x)&=&x^{-0.65}(1-x)^{8.27}(c_{1}+c_{2}\sqrt{x}), \\
\Delta \delta (x)&=&x^{-0.65}(1-x)^{8.27}c_{3}(1+64.9x), \nonumber \\
\Delta G (x)&=&x^{-0.49}(1-x)^{5.51}c_{3}(d_1+d_2 \sqrt{x}+d_3 x). \nonumber 
\end{eqnarray}

We will not consider $\Delta\delta$, the parameter that breaks the isospin 
SU(2) symmetry of a sea, (we assume $\Delta\delta = 0$) because one gets 
that the determination of such parameters is not reliable. It is very 
important what assumptions one makes about the sea contribution. From the 
MRS fit for unpolarised structure functions the natural assumption would be: 
$\Delta \bar{s}=\Delta \bar{d}/2 = \Delta \bar{u}/2$. If we 
add the condition that $a_8$ value should be equal to the value determined 
from the semileptonic hyperon decays, $\Delta s$ is practically determined 
and though also nonstrange sea is determined. For comparison we will also 
consider such model. We assume that $\Delta s$ (strange sea) is described 
by additional parameters namely  
\begin{equation}
\Delta M_i= x^{-0.65}(1-x)^{8.27} (c_{1i}+c_{2i}\sqrt{x})
\end{equation}
where $i = u, d, s$ 
and $c_{1i}$ and $c_{2i}$  for $u$ and $d$ are equal. For the strange 
quarks we have additional independent parameters. Comparing with the 
expression (5) we see that in $\Delta M_i$ there is no term behaving like 
$x^{-1.15}$ at small $x$ (we assume that $\Delta M_i$  and hence all sea 
distributions are integrable) which means that in $\Delta M_i$  coefficient 
in front of this term have to be splitted into equal parts in $\Delta M_i^+$ 
and $\Delta M_i^-$. We see that we have relatively strong singular behaviour 
of $\Delta d_v$, $\Delta M$ and $\Delta G$ for small $x$ values. 
For comparison we will also consider the model in which leading most 
singular terms are put equal to zero namely $b_1=c_{1i}=d_1=0$, that means 
that plus and minus components have the same value for this powers of $x$ 
that means that we investigate the dependence of the model on the leading 
$x$ behaviour of $\Delta q$ and $\Delta G$ (we know from the ref \cite{alt0} 
that such dependence is strong). In the less singular models the dependence 
of calculated parameters in the unobserved region (below $x \leq 0.003$) 
is weak. In the earlier papers we considered the extrapolation of various 
calculated integrals below $x=0.003$ up to 0 assuming Regge type of behaviour 
for small $x$ values. As will be discussed later the less singular models 
give slightly higher $\chi^2$. 

In order to get the unknown parameters in the expressions for polarised quark 
and gluon distributions we calculate the spin asymmetries starting from 
initial $Q^2$ = 1 $\mbox{GeV}^2$ for measured values 
of $Q^2$ and make a fit to the experimental data on spin asymmetries for 
proton, neutron and deuteron targets. The asymmetry $A_1^N(x,Q^2)$ can 
be expressed via the polarised structure function $A_1^N(x,Q^2)$ as
\begin{equation}
A_1^N(x,Q^2)=\frac{ g^{N}_{1}(x,Q^2)}{ F_1^N(x,Q^2)}= 
\frac{ g^{N}_{1}(x,Q^2)}{ F_2^N(x,Q^2)}[2x(1+ R^N(x,Q^2)] 
\end{equation}
\noindent
where  $R^N=(F_2^N-2xF_1^N)/ 2xF_1^N$ and $F_1^N$ and $F_2^N$ are the 
unpolarised structure functions. We will take the experimental $R_N$ from 
the \cite{whit}.
 Polarised structure function $g_1^N(x,Q^2)$ in the next to leading order 
QCD is related to the polarised quark, antiquark and gluon distributions 
$\Delta q(x,Q^2)$, $\Delta \bar{q}(x,Q^2)$, $\Delta G(x,Q^2)$, in the 
following way:
\begin{eqnarray}
g^{N}_{1}(x,Q^2)&=&\frac{1}{2} \sum_{q}e_{q}^2\{\Delta
q(x,Q^2)+\Delta \bar{q}(x,Q^2)+\frac{\alpha_s}{2 \pi}[\delta c_q *(\Delta
q(x,Q^2)  \nonumber \\
&+&\Delta \bar{q}(x,Q^2))+\frac{1}{f}\delta c_g* \Delta G(x,Q^2)]\}
\end{eqnarray}
\noindent
with the convolution * defined by:
\begin{equation}
(C*q)(x,Q^2) = \int_{x}^{1}\frac{dz}{z} C(\frac{x}{z})q(z,Q^2)
\end{equation}

The explicit form of the appropriate spin dependent Wilson coefficient in 
the $\overline{MS}$ scheme can be found for example in ref. \cite{ew}. The 
NLO expressions for the unpolarised (spin averaged) structure function 
$F^N(x,Q^2)$ is similar to the one in eq.(11) with $\Delta q(x,Q^2)
\rightarrow q(x,Q^2)$ and the unpolarised Wilson coefficients are given for 
example in \cite{ewol1,ewol2}.

The $Q^2$ evolution of the parton densities is governed by the DGLAP 
equations \cite{glap}. For calculating the NLO evolution of the spin 
dependent parton distributions $\Delta q(x,Q^2)$ and $\Delta G(x,Q^2)$ and 
spin averaged $q(x,Q^2)$ and $G(x,Q^2)$ we will follow the method described 
in \cite{ewol2,grv}. We will 
calculate Mellin n-th moment of parton distributions $\Delta q(x,Q^2)$ and 
$\Delta G(x,Q^2)$ according to
\begin{equation}
\Delta q^n(Q^2)=\int_0^1 dx x^{n-1} \Delta q(x,Q^2)
\end{equation}
and then use NLO solutions in Mellin n-moment space in order to calculate 
evolution in $Q^2$ for non-singlet and singlet i.e. of 
$\Delta \Sigma^n(Q^2)=\sum_q [\Delta q^{n}(Q^2)+\Delta \bar{q}^{n}(Q^2)]$
and $\Delta G(Q^2)$.

In calculating evolution of $\Sigma(Q^2)$ and $\Delta G(Q^2)$ with 
 $Q^2$ we have 
mixing governed by the anomalous dimension 2x2 matrix. We used explicit 
formulae given in \cite{ewol2}. Having evolved moments one can insert them 
into the n-th moment of eq.(11).
\begin{eqnarray}
g^{n}_{1}(Q^2)&=&\frac{1}{2} \sum_{q}e_{q}^2 \{\Delta
q^n(Q^2)+\Delta \bar{q}^n(Q^2)+\frac{\alpha_s}{2 \pi}[\delta c_{q}^{n} 
\cdot(\Delta q^n(Q^2)  \nonumber \\
&+&\Delta \bar{q}^n(Q^2))+\frac{1}{f}\delta c_{g}^{n}\cdot \Delta G^n(Q^2)]\}
\end{eqnarray}
\noindent
and then numerically Mellin invert the whole expression. In this way we get 
$g_1(x,Q^2)$. The same procedure is applied taking the appropriate formulas 
giving the different $Q^2$ dependence and the correction coefficients for the 
unpolarised structure functions. Having calculated the asymmetries according 
to equation (10) for the measured in experiments value of $Q^2$ we can make a 
fit to a measured asymmetries on proton neutron and deuteron targets. We 
will take into account for proton 7 points of E80 \cite{e80} and 16 points of 
E130 \cite{e130} of SLAC experiments, 10 points of EMC \cite{emc} and 59 
points of SMC \cite{SMCnew} from CERN and 82 points of E143 \cite{e143new} 
from SLAC. For deuteron we have 65 points from SMC \cite{SMCnew} and 82 
points from E143 \cite{e143new} whereas for neutron 33 points of E142 
\cite{e142n} and 11 points of E154 \cite{e154n} experiments from SLAC and 9 
points from DESY Hermes experiment \cite{hermes}. The last two sets of data 
from E154 and Hermes are taken in order to have more data from neutron target 
and to balance huge 
number of points from proton and deuteron targets. All together we have 374 
data points and together with the assumed $g_8 = 0.579 \pm 0.1$ value we 
have 375 points (174 for proton, 147 for deuteron and 53 for neutron). 

We get the following values of parameters from the fit to all existing (above
 mentioned) data for $Q^2 \geq 1 \mbox{GeV}^2$ for spin asymmetries:
\begin{equation}
\begin{array}{lll}
a_{1}=\hspace*{0.35cm} 0.66,&a_{2}=-4.21,&a_{3}=\hspace*{0.13cm} 14.6,\\
b_{1}\hspace*{0.06cm}=-0.02,&b_{2}\hspace*{0.06cm}= -0.84,
&b_{3}\hspace*{0.055cm}=-1.74,\\
c_{1u}=c_{1d}=-0.28,&c_{2u}= c_{2d}= 3.08,\\
c_{1s}\hspace*{0.055cm}=\hspace*{0.00cm} -0.42,&c_{2s}\hspace*{0.055cm}=
-1.15,&c_{3}\hspace*{0.06cm}=0 
\hspace*{0.2cm} \mbox{(input)},\\
d_1=\hspace*{0.35cm} 2.201&d_2=-22.47&d_3=\hspace{0.13cm} 42.20.
\end{array}
\end{equation}
\noindent
The resulting $\chi^{2}$ per degree of freedom is 
$\chi^{2}/N_{DF}$ =$\frac{308.66}{375-13}$ = 0.853.

The obtained quark and gluon distributions lead for 
$Q^2$ =1 $GeV^2$ to the following integrated quantities:
$\Delta u = 0.77$ ($\Delta u_v = 0.70$, $2\Delta \bar{u} = 0.07$), 
$\Delta d= -0.49$ ($\Delta d_v = -0.56$, $2\Delta \bar{d} = 0.07$),   
$\Delta s= - 0.15$. These numbers give the following predictions:
$\Delta \Sigma = 0.13$ , $\Delta M = 0.00$, $\Delta G=0.22$, 
$\Gamma_1^p = 0.113$, $\Gamma_1^n = -0.062$, $\Gamma_1^d = 0.024$, 
$g_A = \Delta u - \Delta d = 1.26$.

We have relatively small positively polarised sea for up and down quarks and 
stronger negatively polarised sea for strange quarks. The gluon polarisation 
is positive but very small. The value of $g_A$ was not assumed as an input 
in the fit and comes out correctly.

As was already stressed in \cite{grv}, the asymptotic behaviour at small $x$ 
of our polarised quark distributions is determined by the unpolarised ones 
and these do not have the expected theoretically Regge type behaviour or pQCD 
which is also used by experimental groups, to extrapolate results to small 
values of $x$. Some of the quantities change rapidly for 
$x \leq 0.003$.

 Now we will present quantities integrated over the region from 
$x$=0.003 to $x$=1 (it is practically integration over the region which is 
covered by the experimental data, except of non controversial extrapolation 
for highest $x$). The corresponding quantities are
$\Delta u = 0.78$ ($\Delta u_v = 0.67$, $2\Delta \bar{u} = 0.11$), 
$\Delta d = -0.42$ ($\Delta d_v = -0.53$, $2\Delta \bar{d} = 0.11$),   
$\Delta s = - 0.12$, $\Delta \Sigma = 0.24$, $\Delta M = 0.11$, 
$\Delta G = 0.06$.
We can also calculate $\Gamma^p$, $\Gamma^n$ and $\Gamma^d$ in the measured 
region for $Q^2$ = 5 $\mbox{GeV}^2$ and compare them with the quantities 
given by the experimental groups. 

We get $\Gamma_1^p =0.119$, $\Gamma_1^n =-0.078$ and $\Gamma_1^d =0.019$ 
in the whole region, whereas in the region between  $x=0.003$ and $x=0.8$ 
(covered by the data) we have $\Gamma_1^p =0.125$, $\Gamma_1^n =-0.051$ 
and $\Gamma_1^d = 0.034$. The experimental group SMC present \cite{smcth} the 
following values in the measured region (for $Q^2= 5$~$\mbox{GeV}^2$):
\begin{eqnarray}
\Gamma_1^p &=& \hspace*{0.33cm} 0.130 \pm 0.007, \nonumber \\
\Gamma_1^n &=& -0.054\pm 0.009,\\
\Gamma_1^d &=& \hspace*{0.33cm} 0.036 \pm 0.005. \nonumber 
\end{eqnarray}

The world average for such $Q^2$ is \cite{smcth} for the whole region 
($0 \leq x \leq 1$):
\begin{eqnarray}
\Gamma_1^p &=&\hspace*{0.33cm} 0.121 \pm 0.018, \nonumber \\
\Gamma_1^n &=& -0.075\pm 0.021, \\
\Gamma_1^d &=& \hspace*{0.33cm} 0.021\pm 0.017. \nonumber 
\end{eqnarray}
\noindent Our results are in good agreement with given experimental values.
For comparison we have also made fits using formulas of the simple parton 
model (as in our papers before \cite{bt1}) neglecting evolution of parton 
densities with $Q^2$. More detailed result of these fits (integrated 
densities and so on) will be given later.
\begin{figure}[hbt]\vskip 5.5cm\relax\noindent\hskip -0.2cm
       \relax{\includegraphics{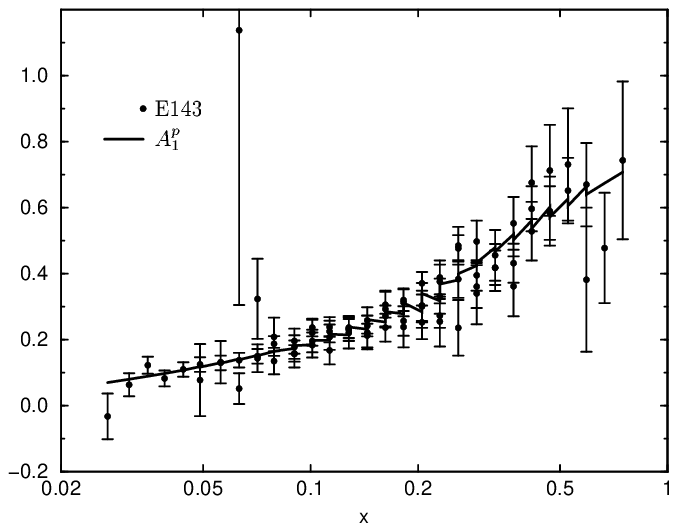}}
        \relax\noindent\hskip 6.9cm
       \relax{\includegraphics{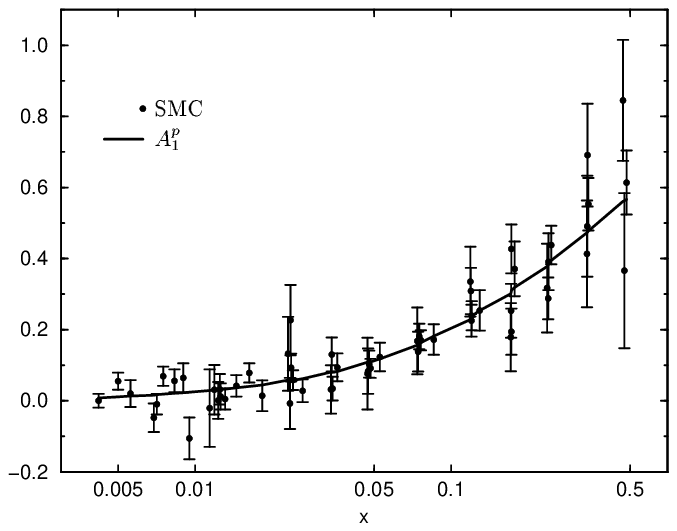}}
\vskip-0.2cm
\label{fig1}
\caption{\footnotesize\em Plots of proton spin asymmetry predicted by our 
basic fit (for experimental~$Q^2$). The data points from different experiments
(E143, SMC) with total errors are also shown.}
\end{figure}

In figs.1, 2 and 3 we present the comparison of our basic fit with 
measured asymmetries for proton (1), deuteron (2) and neutron (3) 
targets. The curves 
are obtained by joining the calculated values of asymmetries corresponding 
to actual values of $x$ and $Q^2$ for measured data points. For the same 
value of $x$ we have experimental points corresponding to different $Q^2$ 
values. We see that distribution of experimental points is much bigger than 
the lengths of vertical lines measuring the changes of influence of evolution 
in $Q^2$ for different values of $Q^2$ and the same value of $x$. It seems 
that with  such errors it is difficult to see $Q^2$ dependence of asymmetries.

\begin{figure}[htb]\vskip 5.5cm\relax\noindent\hskip -0.2cm
       \relax{\includegraphics{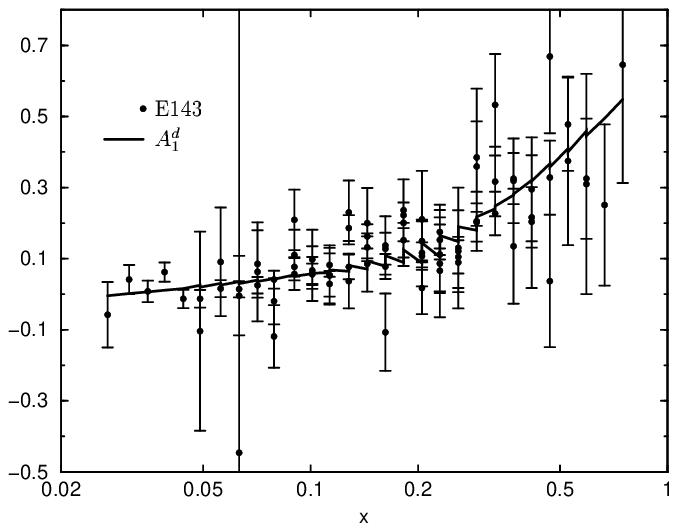}}
        \relax\noindent\hskip 6.9cm
       \relax{\includegraphics{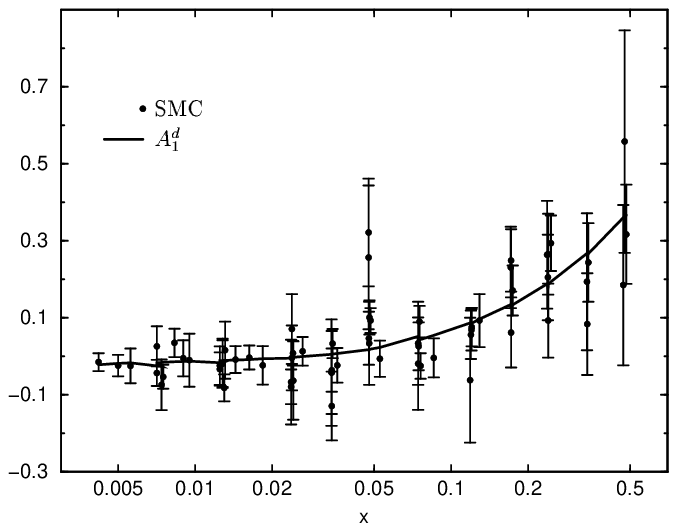}}
\vskip-0.2cm
\label{fig2}
\caption{\footnotesize\em Plots of deuteron spin asymmetry predicted by 
our basic fit (for experimental~$Q^2$). The data points from different 
experiments (E143, SMC) with total errors are also shown.}
\end{figure}
\vspace*{0.5cm}
\begin{figure}[hbt]\vskip 4.2cm\relax\noindent\hskip -0.2cm
       \relax{\includegraphics{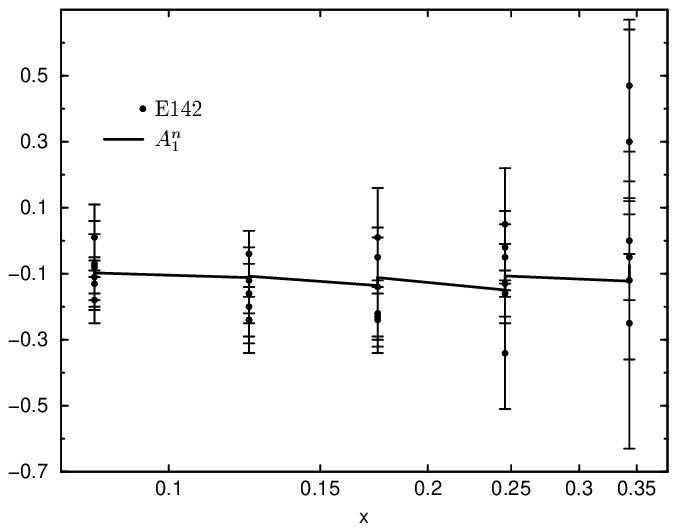}}
        \relax\noindent\hskip 6.9cm
       \relax{\includegraphics{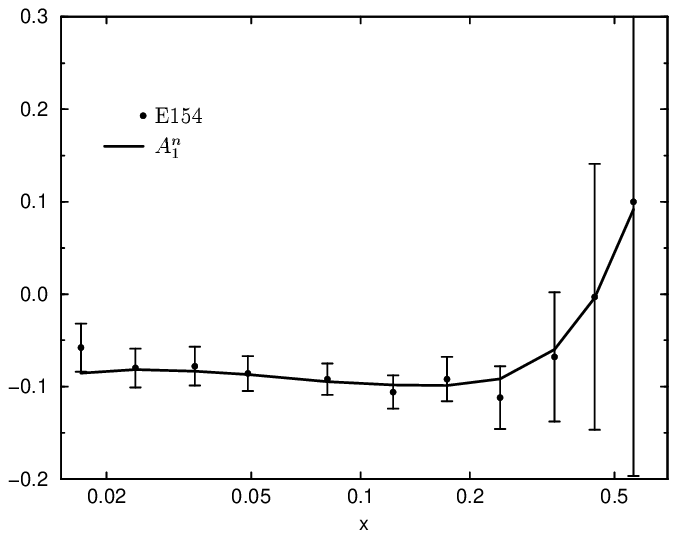}}
\vskip-0.2cm
\label{fig3}
\caption{\footnotesize\em Plots of neutron spin asymmetry predicted by our 
basic fit (for experimental~$Q^2$). The data points from different experiments 
(E142, E154) with total errors are also shown.}
\end{figure}

For asymmetries the curves with $Q^2$ evolution taken into account and 
evolution completely neglected do not differ very much so we do not present 
them. We see that in the case of $g_1$ function for proton (fig.4) the dashed 
line corresponding to the fit with 
no evolution in $Q^2$ taken into account (parton model) that for small $x$ is 
continuation of E143 data (small errors) but lays within experimental errors 
of SMC results. May be it could be considered as some tendency in evidence 
for seeing $Q^2$ dependence in the data but certainly not a very strong one. 
In the deuteron and neutron data the effect is even less pronounced.

\begin{figure}[htb]\vskip 5.7cm\relax\noindent\hskip -0.2cm
       \relax{\includegraphics{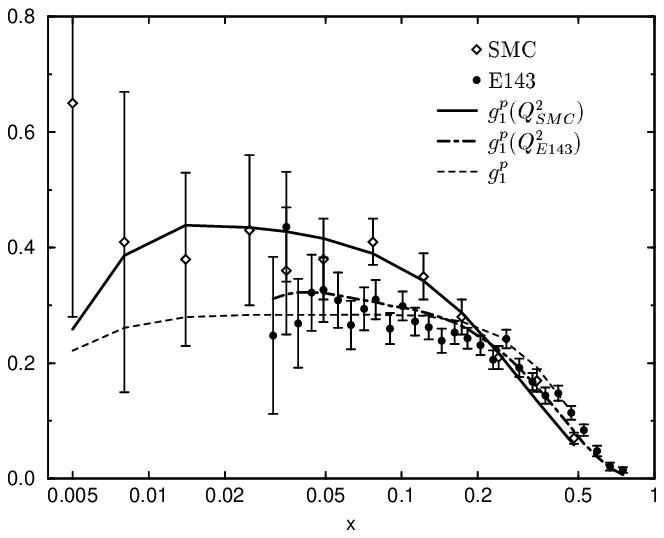}}
        \relax\noindent\hskip 6.9cm
       \relax{\includegraphics{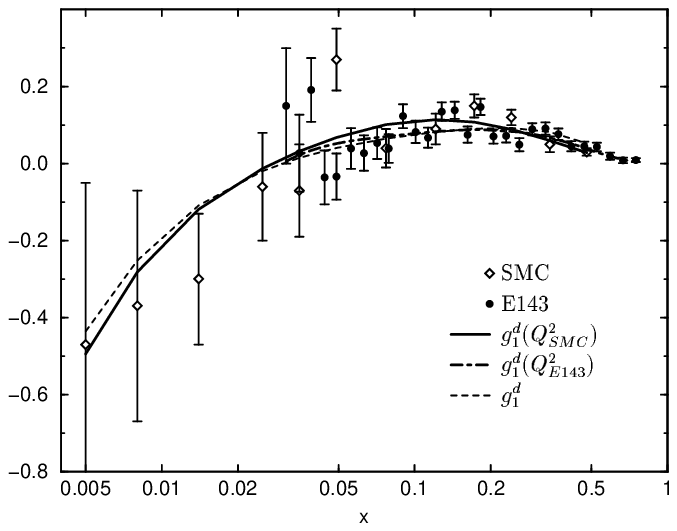}}
\vskip 5.9cm\relax\noindent\hskip 3.4cm
       \relax{\includegraphics{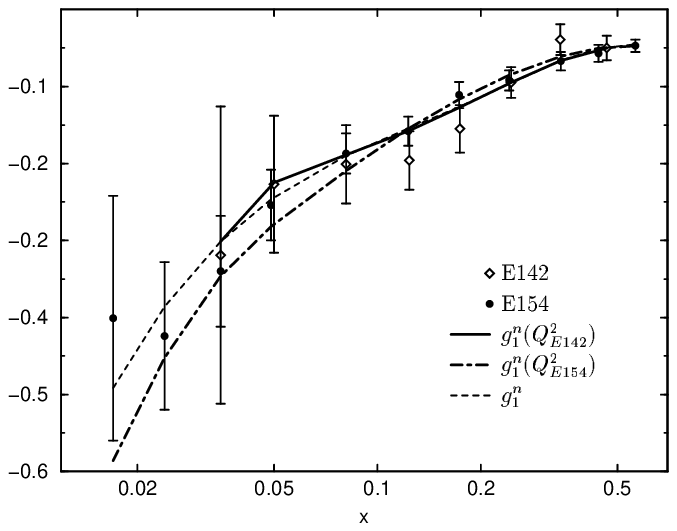}}
\vskip-0.9cm
\label{fig4}
\caption{\footnotesize\em
Plots of structure function $g_1(x,Q^2)$ obtained in our fit 
(for experimental values of $Q^2$) and compared with data points for proton, 
deuteron and neutron targets. The solid curves are predictions for SMC or 
E142 experimental points, whereas dot-dashed ones are for E143 or E154 data. 
The dashed curves are predictions from the fit with no $Q^2$ evolution of 
considered structure functions.}
\end{figure}
Polarised quark distributions for up and down valence quarks as well as non 
strange, 
strange quarks and gluons for $Q^2$ = 1 $\mbox{GeV}^2$ are presented in 
figure 5. They are compared with the distributions when $Q^2$ evolution is not 
taken into account and with the corresponding unpolarised quark 
and gluon distributions. We see that especially in the case of polarised 
gluon distribution function does not resemble the distribution function for 
unpolarised case. This function is also quite different from the gluon 
distribution (given in \cite{gerst}) used to estimate $\Delta G/G$ in COMPASS 
experiment planned at CERN \cite{nassalski}.

\begin{figure}[htbp]\vskip 5.5cm\relax\noindent\hskip -0.2cm
       \relax{\includegraphics{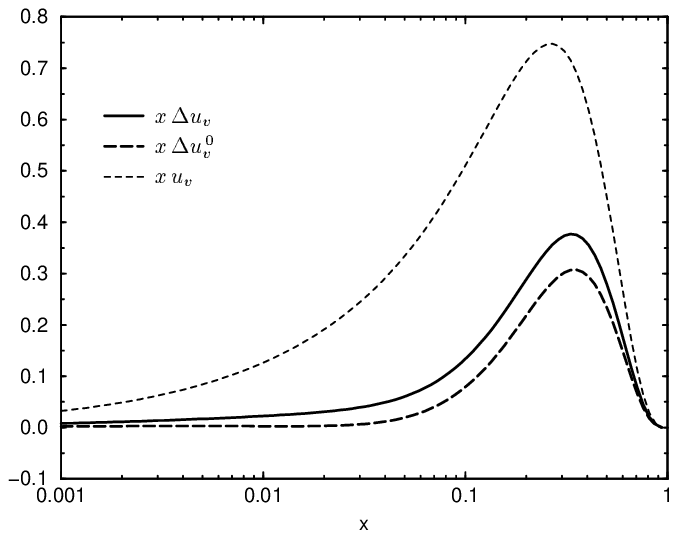}}
        \relax\noindent\hskip 6.9cm
       \relax{\includegraphics{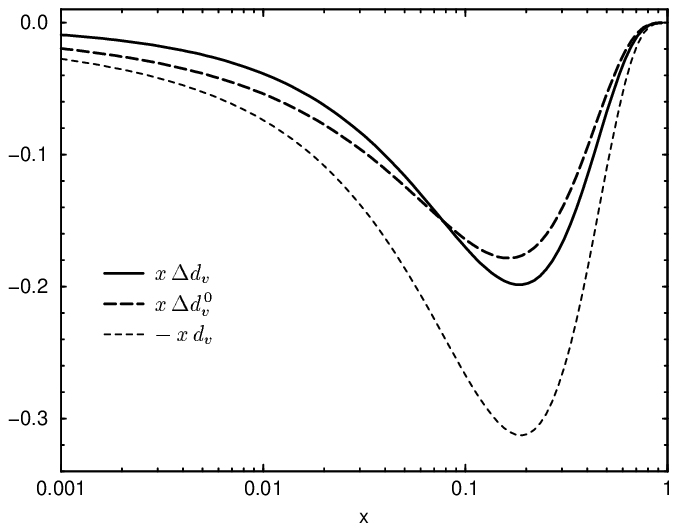}}
\vskip 5.0cm\relax\noindent\hskip -0.2cm
       \relax{\includegraphics{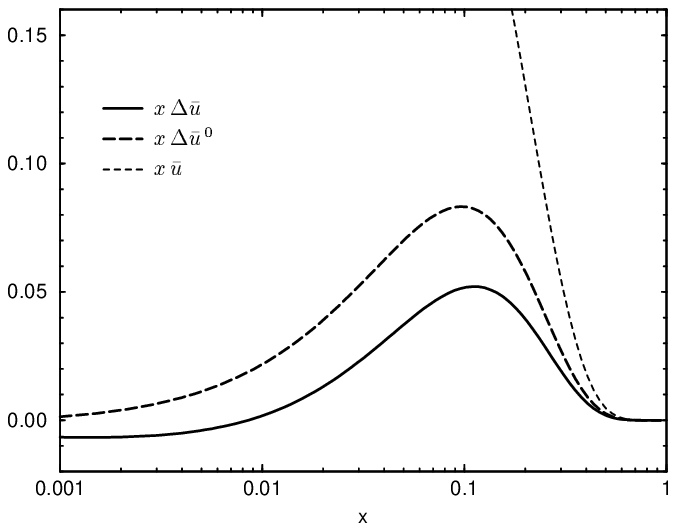}}
        \relax\noindent\hskip 6.9cm
       \relax{\includegraphics{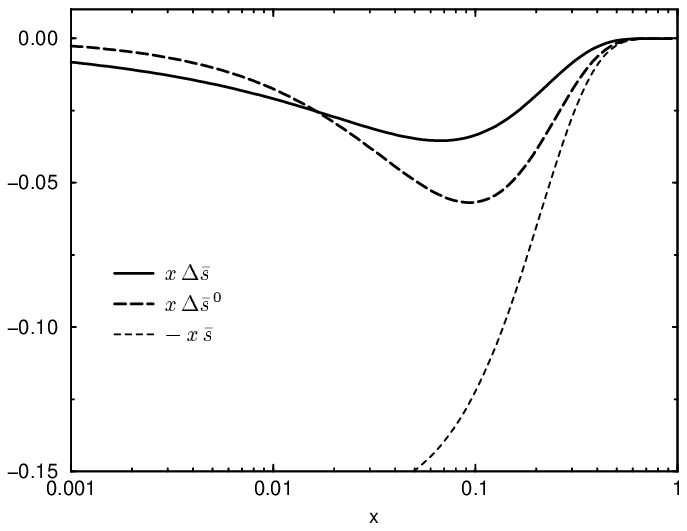}}
\vskip 5.5cm\relax\noindent\hskip 3.4cm
       \relax{\includegraphics{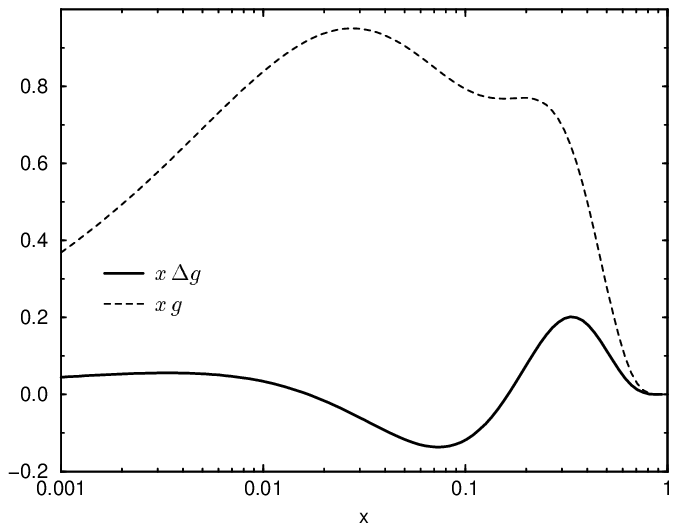}}
\vskip-1.0cm
\label{fig5}
\caption{\footnotesize\em
Polarised quark ($x\Delta u_v$, $x\Delta d_v$), antiquark 
($x\Delta \bar{u}$, $x\Delta \bar{s}$) and gluon distributions 
($x\Delta G$) predicted by our basic fit at $Q^2=1 \mbox{GeV}^2$ (solid 
curve). For comparison prediction for such quantities for the fit without 
$Q^2$ evolution taken into account (long-dashed curves) as well as 
nonpolarised quark, antiquark and gluon distributions from \cite{mrsn} 
(dashed ones) are also shown.}
\end{figure}

Fixing the value of $g_8$ is very important for the fit. When we relax the 
condition for $g_8=0.579$, $\chi^2$ 
goes a little bit down to the value 308.65. We get the fit with the 
parameters not very different from our basic fit but with very small 
$g_8$ (0.03) and bigger $\Delta \Sigma = 0.27$, ($\Delta \Sigma =0.37$ for 
$0.003 \leq x \leq 1)$
 and positive $\Delta s=0.08$, $\Delta \bar{u}=\Delta \bar{d}=0.03$. 
It means that fixing the value of $g_8$ equal to experimental value (gotten 
from hyperon $\beta$-decays data) enforces the negative value of $\Delta s$ 
between -0.15 and -0.12. The obtained solution without fixing $g_8$ value is 
somehow dual to our fit, $g_8$ very small and $\Delta \Sigma$ relatively 
large comparing with $g_8$ close to its experimental value and $\Delta 
\Sigma$ rather small. 

In order to check 
how the fit depends on the assumptions made about the sea contribution we 
have also made fit with $\Delta \bar{u}=\Delta \bar{d}=2\Delta \bar{s}$, the 
assumption that follows directly from MRS unpolarised fit. The $\chi^2$ 
value per degree of freedom $\chi^{2}/N_{DF}$ =$\frac{311.50}{375-11}$ = 
0.856 is a little bit worse. In this case we have
$\Delta u = 0.76$ ($\Delta u_v = 0.98$, $2\Delta \bar{u} = -0.21$), 
$\Delta d = -0.48$ ($\Delta d_v = -0.26$, $2\Delta \bar{d} = -0.21$), 
$\Delta s = - 0.11$, $\Delta \Sigma = 0.18$, $\Delta M=-0.53$, 
$\Delta G = 0.28$. Hence, we see that with the different assumption about 
sea behaviour the overall sea contribution changes quite drastically. The 
quantity $\Delta s$ must be negative in order to get experimental value for 
$g_8$ and because of our assumption ($\Delta \bar{u}=\Delta \bar{d}=2\Delta 
\bar{s}$) we obtain relatively big negative values of non strange sea 
for up and down quarks. We see that the values for sea polarisation depend 
very strongly on the taken assumptions (in many papers \cite{grv,alt0,gerst} 
SU(3) symmetric sea is assumed that also together with fixing of $g_8$ value 
gives negative non strange sea). It means that the sea contribution is not 
very well determined. On the other hand $\Delta u=\Delta u_v+2\Delta \bar{u}$ 
and $\Delta d=\Delta d_v+2\Delta \bar{d}$ practically do not change (however, 
$\Delta u_v$ and $\Delta d_v$ also change). Also $\Delta G$ does not change.

Looking at the 
dependence of unpolarised quark and gluon densities we see 
that the most singular behaviour for small $x$ we have for $d_v(x)$, $M(x)$
and $G(x)$. For comparison we have investigated the model when in 
polarised densities 
these most singular contributions are absent. In this case $\Delta d_v$,
$\Delta M$ and $\Delta G$ are $\sqrt{x}$ less singular than in our basic
fit. For such a fit we get 
$\chi^{2}/N_{DF}$ =$\frac{314.63}{375-9}$ = 0.864 i.e only 
slightly higher than in our basic fit. We get in this case:
$\Delta u = 0.78$ ($\Delta u_v = 0.71$, $2\Delta \bar{u} = 0.07$), 
$\Delta d = -0.41$ ($\Delta d_v = -0.48$, $2\Delta \bar{d} = 0.07$),
$\Delta s = - 0.10$, $\Delta \Sigma = 0.27$, $\Delta M=0.05$, 
$\Delta G = -0.40$. If we do not modify $\Delta G(x,Q^2)$ omitting the most 
singular term $\Delta G$ remains positive.
In this fit integrated quantities taken over the whole range of 
$0 \leq x \leq 1$ and in the truncated one ($0.003 \leq x \leq 1$) differ
very little. $\Delta G$ is negative opposite to the basic fit. So it is 
possible to get the fit of comparable quality to our basic fit with 
practically no change of 
integrated quantities in the region between $x=0$ and $x=0.003$. For 
$Q^2$=1 $\mbox{GeV}^2$ we have $\Gamma_1^p = 0.121$ and 
$\Gamma_1^n = -0.044$ (to be compared with $\Gamma_1^n = -0.062$) 
and relatively big $\Delta \Sigma =0.27$.

The obtained results can 
be compared with the fit when instead of 374 points for different $x$ and 
$Q^2$ values we take spin asymmetries for only 130 data points with the 
averaged values for the same $x$, averaged $Q^2$ and smaller errors. We have 
then for proton target points obtained at CERN (10 from EMC, 12 for SMC) and 
at SLAC (4 from E80, 8 from E130 and 28 for E143). For deuteron we take into 
account 12 points from SMC and 28 from E143 whereas for neutron target data 
points from SLAC (8 from E142 and 11 from E154) and DESY(9 from Hermes).
 In this fit first of all the errors are smaller than in our basic fit and 
the ratio of number of neutron to number of deuteron and proton data points 
are different. It seems that the influence of neutron points is stronger 
than in basic fit $\chi^{2}/N_{DF}$ =$\frac{99.55}{131-13}$ = 0.844 is a 
little bit better than in our basic fit. The integrated values for quark
 and gluon densities are: 
$\Delta u = 0.80$ ($\Delta u_v = 0.57$, $2\Delta \bar{u} = 0.23$), 
$\Delta d = -0.54$ ($\Delta d_v = -0.78$, $2\Delta \bar{d} = 0.23$), 
$\Delta s = - 0.16$, $\Delta \Sigma = 0.10$, $\Delta M = 0.31$, 
$\Delta G = -0.69$ and $g_A=1.35$.
We see that averaging over $x$ and $Q^2$ and different numbers of data points 
leads to not a very different fit (for example $\chi^2=101.2$ when the 
parameters of the basic fit are used) but the values for integrated valence 
densities and nonstrage sea contribution are different ($\Delta u$ and 
$\Delta d$ do not differ very much and the difference is smaller for 
integrated quantities in the region 
$0.003 \leq x \leq 1$). Integrated gluon density is negative. The fits become
more similar when in the average fit we fix the $g_A$ value, i.e. the 
condition for valence contribution.

The fact that the averaged and non averaged samples of data 
points results for valence quark densities and sea contributions are 
different and in the case of integrated gluon density even sign is 
different means that these quantities are not very precisely determined 
in the fits. After assuming $g_8$ value from experiment and fixing $\Delta s$ 
contribution we can determine $\Delta u$ and $\Delta d$ values (the 
splitting of $\Delta q$ in $\Delta q_v$ and $2\Delta \bar{q}$ depends on data 
sample and assumptions about sea contribution). It could be that 
differences that come out in comparing fits to the averaged and non averaged 
data are connected with the fact that rather singular 
polarised parton distributions are able to pick up differences in two 
experimental data samples, due to different number of neutron to proton 
and deuteron data points. The fit with less singular densities ($\Delta d_v$, 
$\Delta \bar{u}$, $\Delta \bar{d}$ and $\Delta G$) with $\chi^2$ a little bit 
higher 104.87 ($\chi^{2}/N_{DF}$ =$\frac{104.87}{131-9}$)=0.860) is for
 averaged data points nearly identical to that for non 
averaged one. $\Delta G$ is like in previous case small and negative
 $\Delta G = -0.4$.

As was already mentioned before we have also made for 
comparison fits neglecting evolution of parton densities 
with $Q^2$ (formulas from the simple parton model). We get for non 
averaged data sample 
$\chi^{2}/N_{DF}$ =$\frac{318.25}{375-10}$)= 0.872 (higher than in our 
basic fit where: $\chi^{2}/N_{DF}$ = 0.853):
$\Delta u = 0.68$ ($\Delta u_v = 0.46$, $2\Delta \bar{u} = 0.22$), 
$\Delta d = -0.41$ ($\Delta d_v = -0.63$, $2\Delta \bar{d} = 0.22$),   
$\Delta s = - 0.16$, $\Delta \Sigma = 0.11$, $\Delta M = 0.27$, 
$\Gamma_1^p = 0.120$, $\Gamma_1^n$ = -0.062. For averaged data points we get 
$\chi^{2}/N_{DF}$ =$\frac{103.79}{131-10}$)= 0.858 
(this number should be compared  with $\chi^{2}/N_{DF}$ = 0.844, 
the corresponding quantity from NLO fit) and we have:
$\Delta u = 0.69$ ($\Delta u_v = 0.40$, $2\Delta \bar{u} = 0.29$), 
$\Delta d = -0.42$ ($\Delta d_v = -0.72$, $2\Delta \bar{d} = 0.29$),   
$\Delta s = - 0.16$, $\Delta \Sigma = 0.11$, $\Delta M = 0.42$. 
Hence, $\chi^2$ is smaller in the case of averaged sample. We see that also 
in this case the value $\Delta u$ and $\Delta d$ and $\Delta s$ are 
practically the same and there are some shifts in valence values and non 
strange sea contribution (the differences are smaller in the 
$0.003 \leq x \leq 1$ region). Similarly to the case with evolution taken 
into account the fits are nearly identical when we make them with less 
singular $\Delta d_v$, $\Delta \bar{u} = \Delta \bar{d}$ and $\Delta G$ 
densities ($\chi^2$ is in this case higher than in the basic fit for non 
averaged data sample or for averaged one).

It has been pointed out \cite{alt0} that the positivity conditions could be 
restrictive and influence the contribution of polarised gluons. We have also 
made a fit to experimental data without such assumption for polarised 
partons. The $\chi ^2$ value does not changed much 
$\chi^{2}/N_{DF}$ =$\frac{308.32}{375-13}$ = 0.852 and we get 
$\Delta u = 0.79$ ($\Delta u_v = 0.35$, $2\Delta \bar{u} = 0.44$), 
$\Delta d = -0.50$ ($\Delta d_v = -0.94$, $2\Delta \bar{d} = 0.44$),   
$\Delta s = - 0.12$, $\Delta \Sigma = 0.17$, $\Delta M = 0.76$, 
$\Delta G = 0.12$. Hence, sea contribution for non strange quarks became 
big and there are shifts between valence and sea contribution. The gluon 
contribution does not change staying close to zero for non averaged data and 
small and negative for averaged data. The most important restriction is for 
the valence $d$ quark (in the case without positivity assumption it becomes 
big close to -1). Relaxing the positivity of other quantities does not 
change the fit. It seems that 
without positivity the values of parton distributions are strongly pushed in 
the direction of small contribution of valence $u$ quark and big (and 
negative) valence $d$ quark. The gluon contribution is not modified very 
much. Hence, we decided to use the positivity assumptions in our fit.

In summary we have made fits to two samples of data with averaged $x$ and 
$Q^2$ values and nonaverged ones (adding averaged neutron data from E154 
and Hermes experiments). To check the influence of different model dependent 
assumptions we consider fits without fixing $g_8$ value, with modified sea 
contribution and less singular behaviour for valence $d$ quark, sea 
contribution and gluon densities. For comparison we have also considered 
fits to the simple parton model neglecting $Q^2$ dependence of parton 
densities. It seems that splitting of integrated densities $\Delta u$, 
$\Delta d$ in valence and 
sea contribution  is model dependent ($\Delta u$ and $\Delta d$ were more or 
less the same). The integrated gluon contribution comes out relatively small 
and of different sign in averaged and nonaveraged data sample. The 
comparison of $g_1$ for fits without evolution and with $Q^2$ evolution of 
parton densities taken into account is given. It seems that experimental 
accuracy is still not enough to make precise statements about polarised 
quark and gluon densities.

\end{document}